\documentstyle[preprint,aps,prd]{revtex}

\begin{document}
\title{Creation of unstable particles and decoherence in semiclassical 
cosmology}
\author{Mario Castagnino \footnote{Electronic address: castagni@iafe.uba.ar}}

\address{{\it Instituto de Astronom\'\i a y F\'\i sica del Espacio - IAFE}\\
 Casilla de Correo
67, Sucursal 28, \\ 1428 Buenos Aires, Argentina}
\author{Susana Landau \footnote{Electronic address: 
slandau@natura.fcaglp.unlp.edu.ar}}

\address{Facultad de Astronom\'\i a y Geof\'\i sica\\
Universidad Nacional de La Plata - La Plata, Buenos Aires - Argentina}

\author{Fernando C.\ Lombardo \footnote{Electronic address:
lombardo@df.uba.ar}}

\address{{\it
Departamento de F\'\i sica, Facultad de Ciencias Exactas y Naturales\\
Universidad de Buenos Aires - Ciudad Universitaria,
Pabell\' on I\\
1428 Buenos Aires, Argentina}}

\maketitle

\begin{abstract}
We consider a simple cosmological model in order to show 
the importance of unstable particle creation for the validity of the 
semiclassical approximation. Using 
the mathematical 
structure of rigged Hilbert spaces we show that particle creation is the 
seed of decoherence which enables the quantum to classical transition. 

\end{abstract}

\maketitle
\title{}
\author{}

\section{Introduction}

We know that the laws of classical mechanics describe with a high 
degree of
accuracy the behavior of macroscopic systems. And yet, it is 
believed that
phenomena on all scales, including the entire Universe, 
follow the
laws of quantum mechanics. So, if we want to reconcile our two last
statements, it is essential to understand the transition from the 
quantum to
the classical regime. One of the scenarios where this problem is 
relevant is
quantum cosmology, in which one attempts to apply quantum 
mechanics to
cosmology. This involves a problem that has not been solved; 
namely,
quantizing the gravitational field. Therefore as a first attempt, it is 
an important issue to
predict the conditions under which the gravitational field may be 
regarded
as classical.

The quantum to classical transition is a very old and interesting 
problem relevant in many branches of physics. It involves the 
concepts of {\it correlations}, i.e., the Wigner function of the 
quantum system should have a peak at the classical trajectories \cite{9}, 
and 
{\it decoherence}, that is, there should be no interference between 
classical trajectories \cite{6}. The density matrix should be 
approximately diagonal. 
In order to understand the emergence of classical 
behaviour, it is esencial to consider the interaction between system and 
environment, since both the decoherence process and the onset 
of classical correlations depend strongly on this interaction.
Both ingredients are not independent and excess of decoherence can destroy 
the correlations \cite{8if}.

In a previous work \cite{1}, one of us has studied the problem of 
choosing an alternative 
mathematical structure, based on a new spectral decomposition with generalized 
 unstable states, which is useful to explain time asymmetry of 
differents models. Following \cite{1}, we will show that this unstable 
quantum states 
satisfy correlation conditions and also produces decoherence 
between different cosmological branches. 
>From this work, we know that if we want to retain the time-
symmetric laws of
nature and at the same time explain the time asymmetry of the 
universe, we
must choose a space of solutions which is not time-symmetric. A 
convenient
choices of time-asymmetric spaces was already proposed in 
Ref. \cite{casta2}.

The scheme is based in the existence of a physically admissible quantum 
superspace $\Phi_-$ and therefore also the existence of a superspace of 
time inverted states of $\Phi_-$, namely  a physically forbidden  
quantum superspace 
$\Phi_+$. Thus, the time invertion that goes from $\Phi_-$ to $\Phi_+$ is 
also forbidden \cite{1}. If the generalized states in $\Phi_-$ are restricted 
to be included in the superspace of regular states ${\cal S}$ (and the same 
for $\Phi_+$ with ${\cal S}^{\times}$ where ${\cal S}^{\times}$ is the space 
of an (anti)linear functional over ${\cal S}$), our real mathematical 
structure is the Gel'fand triplet (or rigged hilbert space) \cite{1}:

\begin{equation} {\cal S} \subset {\cal H}\subset {\cal S}^{\times}.
\end{equation}
 If $K$ is the Wigner time-reversal 
operator we have

\begin{equation} K : \Phi_- \rightarrow  \Phi_+  ~~~ ; ~~~ K:
\Phi_+ \rightarrow \Phi_-.\label{k1}
\end{equation}
Using these spaces of ``generalized" states we can 
also find time-asymmetry for the generalized states. If we choose $\Phi_-$ as 
in Ref. \cite{1}, Eq. (\ref{k1}) means that these generalized states will 
be (growing or decaying) Gamow vectors. Decaying states are transformed into 
growing states (or vice-versa) by time-invers\-ion. 

As we have said \cite{1}, the choice of $\Phi_-$
(or $\Phi_+$) as our space of quantum states implies that $K$ is not
defined inside 
$\Phi_-$ (or $\Phi_+$), so that time-asymmetry 
naturally appears. 
 
But, in the cosmological case, the choice between $\Phi_-$ or $\Phi_+$ (or 
between the periods $t>0$ or $t<0$, or between the two semigroups) is 
conventional and irrelevant, since these objetcs are identical (namely one 
can be obtained 
from the other by a mathematical transformation), and therefore the 
universes, that we will obtain with one choice or the other, are also 
identical and not distinguishable. Only the names ${\it past}$ and {\it 
future} or ${\it decaying}$ and ${\it growing}$ will change but physics is 
the same, i.e., we will always have equilibrium, decoherence, 
growing of entropy, etc. toward, what we would call 
the future. But once the choice is made, a substantial difference is 
established in the
model: using $\Phi_-$ it can be proved that the time evolution operator is 
just $U(t)=e^{-iHt}$, $t>0$, and
cannot be inverted (if the choice would be $\Phi_+$ the condition would change 
 to $t<0$). Therefore even if we continue using the same reversible 
evolution equations, 
the choice of $\Phi_-$ (or which is the same 
$\Phi_+$) introduces time-asymmetry, since now we 
are working in a space where future is substantially different than past. Thus
 the arrow of time is not put {\it by hand} since the choice between the 
period $t>0$ and $t<0$ or between $
\Phi _{-}$ and $\Phi _{+}$ is trivial and unimportant (namely to chose the 
period $t>0$ as the physical period and consider $t<0$ as non-existent, 
because the period before the ``creation of the Universe" is physically 
unaccesible to us or viceversa). The important choice
is between ${\cal H}$ (the usual Hilbert space) and $\Phi _{-}$ 
(or $\Phi _{+})$ as
the space of our physical states. And we are free to make this choice, since
a good physical theory begins by the choice of the best mathematical
structure that mimic nature in the most accurate way.

As far as we know the new formalism is mathematically rigorous and the physical
results of both ones are the same. Two of us have shown this method applied 
to a semiclassical Robertson-Walker metric coupled to a quantum field 
\cite{5}. In this article we have shown how to implement this formalism 
in a semiclassical cosmological model in order to prove tha validity 
of the semiclassical approximation. Decoherence 
and correlations are two necesary ingredients to obtain classical behaviour. 
In Ref. \cite{5} we have proved that the model satisfies both requirements for
classicality. However, paper \cite{5} was the first step to prove 
our mathematical structure in a simple cosmological model; we can rise two 
relevant observations about the validity of the semiclassical approximation:

1) considering the infinite set of unstable modes leads to perfect 
decoherence, destroying correlations\cite{6,7}, as we will prove here.

2) the existence of correlations was proved for only one mode of the scalar 
field and not for the entire density matrix.

In the present article we complete and improve our previous work in order 
to obtain the semiclassical limit as a consequence of the real ``balance'' 
between decoherence and correlations.

In the context of semiclassical cosmology from a fully quantized
cosmological model, the cosmological  scale factor can be defined as 
$a=a\left(\eta\right) $, with $\eta$ the conformal time. 
When $\eta \rightarrow \infty $ we will obtain a classical
geometry g$_{\mu \nu }^{out}$ for the Universe. In the semiclassical 
point of view, the Wheeler-De-Witt equation splits in a classical 
equation for the spacetime metric and in a 
Schr\"{o}dringer equation for the scalar field modes, 
with the corresponding hamiltonian $h\left( a_{out}\right) $. Using 
$h\left(
a_{out}\right) $ and the classical geometry g$_{\mu \nu }^{out}$ we 
can find
a semiclassical vacuum state $\left| 0,out\right\rangle $ which 
diagonalizes
the hamiltonian; and the creation and annihilation operators related to 
this vacuum and the corresponding Fock spaces.

In this paper, we choose time-asymmetric Fock spaces to study a 
simple
cosmological model; we analyze how this model fulfills the two 
requirements for classicality.

The organization of this paper in the following. In section II we 
introduce
the cosmological model and we summarize our previous results of Refs. \cite{1} 
and  \cite{5}. In section III we analyze the
conditions for the existence of decoherence and correlations in this 
model.
Since we achieve perfect decoherence, in Section IV we need to introduce a 
cutoff. We
suggest a particular value for the cutoff using a relevant 
physical
scale that ensures the validity of the semiclassical approximation, namely 
the Planck scale. In section V we briefly discuss our results.

\section{The model and previous results}

In this Section we will only extract the main results of Ref. \cite{5}. 
Let us consider a flat Robertson-Walker spacetime coupled to a massive 
conformally coupled scalar field. In the specific model of \cite{5} we 
have considered a graviatational action given by
 
\begin{equation}
S_g=M^2\int d\eta \,\left[ -%
{\textstyle {1 \over 2}}
\stackrel{.}{a}^2-V\left( a\right) \right],  \label{accion grav}
\end{equation}
where $M$ is Planck's mass, $\stackrel{.}{a}\ =\frac{da}{d\eta }$ 
and $V(a)$
is the potential function that arises from a spatial curvature, a 
possible
cosmological constant and, eventually a classical matter field.

In this paper we will consider the potential function used by Birrell 
and
Davies \cite{2} to illustrate the use of the adiabatic approximation in 
an
asymptotically non-static four dimensional cosmological model:

\begin{equation}
V\left( a\right) =\frac{B^2}2\left( 1-\frac{A^2}{a^2}\right),  \label{pot}
\end{equation}
where A and B are arbitrary constants. 

The Wheeler-DeWitt equation for this model is:

\begin{equation}
H\Psi \left( a,\varphi \right) =\left( h_g+h_f+h_i\right) \Psi \left(
a,\varphi \right) =0,  \label{h}
\end{equation}
where

\begin{equation}
h_g=\frac 1{2M}\partial _a^2+M^2V\left( a\right),  \label{h1}
\end{equation}

\begin{equation}
h_f=-%
{\textstyle {1 \over 2}}
\int_k\left( \partial _{\varphi _k}^2-k^2\varphi _k^2\right) dk,  \label{h2}
\end{equation}

\begin{equation}
h_i=\frac{m^2a^2}2\int_k\varphi _k^2dk,  \label{h3}
\end{equation}
and $m$ is the mass of the scalar field.

In the semiclassical approximation, where the geometry is 
considered as classical, and only the scalar field is quantized, we 
propose a WKB solution to the Wheeler-DeWitt equation:

\begin{equation}
\Psi \left( a,\varphi \right) =\chi \left( a,\varphi \right) \exp \left[
iM^2S\left( a\right) \right],
\end{equation} where $S$ is the classical action for the geometry.

To leading order (i.e. $M^2$), we get:

\begin{equation}
\left[ \frac{dS\left( a\right) }{da}\right] ^2=2V\left( a\right),  \label{v1}
\end{equation}
which is essentially the Hamilton-Jacobi equation for the variable 
$a\left(
\eta \right) $. Fron this equation we can find the classical solutions

\begin{equation}
a\left( \eta \right) =\pm \left( A^2+B^2\eta ^2\right) ^{%
{\textstyle {1 \over 2}}
}+C,  \label{potencial}
\end{equation}
where C is a constant.

Taking the following order in the WDW equation, we obtain a 
Schr\"{o}dringer
equation for $\chi \left( a,\varphi \right) :$

\begin{equation}
i\frac d{d\eta }\chi \left( a,\varphi \right) =-%
{\textstyle {1 \over 2}}
\int_k\left[ \partial _k^2-\Omega _k^2\varphi _k^2\right] dk\chi \left(
a,\varphi \right),  \label{hamil}
\end{equation}
where $\Omega _k^2=m^2a^2+k^2$

Since the coupling is conformal we will have well-defined vacua \cite{2}. 
So, we
consider now two scales $a_{in}$ and $a_{out}$ such that 
$0<a_{in}<<a_{out}$%
. Next, we define the corresponding $\left| 0,in\right\rangle ,\left|
0,out\right\rangle $ vacua there, where $\left| 0,in\right\rangle $ is 
the
adiabatic vacuum for $\eta \rightarrow -\infty $ and $\left|
0,out\right\rangle $ is the corresponding for $\eta \rightarrow +\infty 
$.
It is well known \cite{2,3} that, in the 
case we are considering, we can diagonalize the time-dependent
 Hamiltonian (Eq. (\ref{hamil})) at $a_{in}$ and a$_{out}$, define the
corresponding creation and annihilation operators, and the 
corresponding Fock spaces.

Thus, following Eqs. $\left[ 37-43\right] $ from Ref. \cite{1} we can
construct the Fock space and find the eigenvector of $h\left( 
a_{out}\right)
,$ as follows:

\begin{equation}
h\left( a_{out}\right) \left| \left\{ k\right\} ,out\right\rangle =h\left(
a_{out}\right) \left| \varpi ,\left[ k\right] ,out\right\rangle =\Omega
\left( a_{out}\right) \left| \left\{ k\right\} ,out\right\rangle
=\sum_{k\varepsilon \left\{ k\right\} }\Omega _\varpi \left( 
a_{out}\right)
\left| \varpi ,\left[ k\right] ,out\right\rangle
,\end{equation}
where $\left[ k\right] $ is the remaining set of labels necessary to 
define
the vector unambiguously and $\left| \varpi ,\left[ k\right]
,out\right\rangle $ is an ortonomal basis \cite{1}.

In the same way we can find the eigenvectors of $h\left( a_{in}\right) 
$.
Thus we can also define the S matrix between the in and out 
states $\left( 
\text{Eq. 44 of Ref. \cite{1}}\right) $:

\begin{equation}
S_{\varpi ,\left[ k\right] ;\varpi ^{\prime },\left[ k^{\prime }\right]
}=\left\langle \varpi ,\left[ k\right] ,in\right| \varpi ^{\prime },\left[
k^{\prime }\right] ,out\rangle =S_{\varpi ,\left[ k\right] ;\left[ k^{\prime
}\right] }\,\delta \left( \varpi -\varpi ^{\prime }\right)
\end{equation}

As we have explained in the Introduction, we will choose time-asymmetric
spaces in order to get a better description of time asymmetry of the
universe. Therefore we make the following choice: for the in Fock 
space we
will use functions $\left| \psi \right\rangle \in \Phi _{+,in}$ namely, 
such
that $\left\langle \varpi ,in\right| \psi \rangle \in S\mid _{R_{+}}$ and 
$%
\left\langle \varpi ,in\right| \psi \rangle \in H_{+}^2\mid 
_{R_{+}}$ where $%
H_{+}^2$ is the space of Hardy class functions from above;\ and for 
the out
Fock space we will use functions $\left| \varphi \right\rangle \in \Phi
_{-,out}$ such that $\left\langle \varpi ,out\right| \varphi \rangle \in
S\mid _{R_{+}}$and $\left\langle \varpi ,out\right| \varphi \rangle \in
H_{-}^2\mid _{R_{+}}.$ So we can obtain a spectral 
decomposition
for the $h\left( a_{out}\right) $ (in a weak sense) \cite{1,5}:

\begin{equation}
h\left( a_{out}\right) =\sum_n\Omega _n\left| \bar{n}\right\rangle
\left\langle \bar{n}\right| +\int dz\,\Omega _z\left| \bar{z}\right\rangle
\left\langle \bar{z}\right|,  \label{ham2}
\end{equation}
where $\Omega _n=m^2a^2+z_n$ and $z_n$ are the poles of the S 
matrix.

>From references \cite{1} and \cite{5} it can be seen that S matrix
corresponding to this model has infinite poles and the mode $k$, 
corresponding to each pole reads:

\begin{equation}
k^2=mB\,\left[ -\frac{m\,A^2}B-2i\,\left( n+%
{\textstyle {1 \over 2}}
\right) \right]\label{k2}.
\end{equation}
Thus we can compute the squared energy of each pole:

\begin{equation}
\Omega _n^2=m^2a^2+mB\,\left[ -\frac{m\,A^2}B-2i\,\left( n+%
{\textstyle {1 \over 2}}
\right) \right].  \label{energia compleja}
\end{equation}

The mean life of each pole is:

\begin{equation}
\tau _n=%
{\textstyle {\sqrt{2} \over 2}}
\frac{\left[ m^2\left( a_{out}^2-A^2\right) +\left( m^4\left(
a_{out}^2-A^2\right) ^2+4m^2B^2\left( n+%
{\textstyle {1 \over 2}}
\right) ^2\right) ^{\frac 12}\right] ^{\frac 12}}{%
\mathop{\rm Im}
\,\ B\ \left( n+%
{\textstyle {1 \over 2}}
\right) }.  \label{vida media}
\end{equation}

Using the spectral 
decomposition (\ref{ham2}) we will show, in the next section, how decoherence 
produces the 
elimination of all quantum interference effects. But we must notice that we 
can introduce this spectral 
decomposition only using the unstable ideal states.

We believe that our results can be 
generalized to other models, since essencially they are based in the 
existence of an infinite set of poles in the scattering matrix. Nevertheless 
the model considered in this paper will allow us to complete 
all the calculations, being therefore a good example of what can be done 
with our method. 

\section{Perfect decoherence and no correlations}

In this section we will show how the complete set of unstable modes 
destroy quantum interference, but also demolish classical correlations.
The appearence of decoherence coming from the spectral decomposition of 
Eq. $\left( 
\text{%
\ref{ham2}}\right) $ shows the importance of the unstable modes in the 
quantum to classical process. It has been proved \cite{12} 
that decoherence 
is closely related to another dissipative process, namely, particle 
creation
from the gravitational field during universe expansion. In Eq. $\left( 
\text{%
\ref{ham2}}\right) $ we obtain as in \cite{5} a set of discrete unstable
states, namely, the unstable particles, and a set of continuous 
stable
states (see Eq. (\ref{ham2})), the latter corresponding to the stable
particles.

As the modes do not interact between themselves we can write: 
\begin{equation}
\chi \left( a,\varphi \right) =\prod_{n=1}^\infty \chi _n\left( \eta
,\varphi _n\right),
\end{equation}
the Schr\"{o}dringer equation for each mode is

\begin{equation}
i\frac d{d\eta }\chi _n\left( a,\varphi _n\right) =-%
{\textstyle {1 \over 2}}
\left[ \partial _n^2-\Omega _n^2\varphi _n^2\right] \ \chi _n\left(
a,\varphi _n\right).  \label{s}
\end{equation}

As usual, we now assume the gaussian ansatz for $\chi _n\left( \eta 
,\varphi
_n\right) :$

\begin{equation}
\chi _n\left( \eta ,\varphi _n\right) =A_n\left( \eta \right) \,\exp \left[
i\,\alpha _n\left( \eta \right) -B_n\left( \eta \right) \,\varphi _n^2\right]
,\label{agaussiano}
\end{equation}
where $A_n\left( \eta \right) $ and $\alpha _n\left( \eta \right) $ are
real, while $B_n\left( \eta \right) $ may be complex, namely, 
$B_n\left(
\eta \right) =B_{nR}\left( \eta \right) +i\,B_{ni}\left( \eta \right) .$

After integration of the scalar field modes, we can define the reduced density 
matrix $\rho _r as:$

\begin{equation}
\rho _r^{\alpha \beta }\left( a,a^{\prime }\right) =\prod_{n=1}^\infty \rho
_{rn}^{\alpha \beta }\left( \eta ,\eta ^{\prime }\right) 
=\prod_{n=1}^\infty
\int d\varphi _n\,\chi _n^\alpha \left( \eta ,\varphi _n\right) \ \chi
_n^\beta \left( \eta ,\varphi _n\right).  \label{matriz2}
\end{equation} where $\alpha $ and $\beta $ symbolizes the two different 
classical geometries.

It is convenient to introduce the following change of variable in order 
to characterize the wave function of each mode:

\begin{equation}
B_m=-%
{\textstyle {1 \over 2}}
\ \frac{\dot{g}_m}{g_m}.
\end{equation}

where $g_N$ is the wave function that represents the quantum state of the
universe being also the solution of
the differential equation

\begin{equation}\ddot g_m+\Omega_m^2 g_m=0,\end{equation}
$\Omega_m$ can be the complex energy $\Omega_n$ in our treatment. 

In the more general case we use an arbitrary initial state $\vert 0,0\rangle$,
instead of $\vert 0,in\rangle$. From the discussion presented in the 
Introduction, and from Ref. \cite{gamow} we know that,
in a generic case, an infinite set of complex poles does exist. Then we must
change (\ref{k2}) by $k^2=k_n^2$ ($n=0, 1, 2, .....$), where these are the 
points where the infinite poles are located in the complex plane $k^2$; 
thus, $\Omega_n^2$  now reads as

\begin{equation}\Omega_n^2=m^2a^2+k_n^2.\end{equation}
 
We will consider the asymptotic (or adiabatic) expansion of function $g_N$ 
when $a\rightarrow +\infty$ in the basis of the out modes. $g_N$ is the wave 
function
that represents the state of the universe, corresponding to the arbitrary 
initial state; its expansion reads 

\begin{equation}g_m=\frac{P_m}{\sqrt{2\Omega_m}}\exp [-i\int_0^\eta \Omega_m
d\eta]+\frac{Q_m}{\sqrt{2\Omega_m}} \exp [i \int_0^\eta \Omega_m
d\eta],\label{gN}\end{equation}
where $P_m$ and $Q_m$ are arbitrary coefficients showing that $\vert
0,0\rangle$ is really arbitrary.

It is obvious that if all the $\Omega_m$ are real, like in the case of the
$\Omega_k$, (\ref{gN}) will have an oscillatory nature, as well as its 
derivative.
This will also be the behaviour of $B_k$. Therefore the limit of $B_k$
when
$\eta \rightarrow +\infty$ will be not well defined even if $B_k$ itself is
bounded.

But if $\Omega_m$ is complex the second term of (\ref{gN}) will have a damping
factor and the first a growing one. In fact, the complex extension of Eq.
(\ref{gN}) (with $m=n$) reads

\begin{equation}g_n=\frac{P_n}{\sqrt{2\Omega_n}}\exp [-i\int_0^\eta \Omega_n
d\eta]+\frac{Q_n}{\sqrt{2\Omega_n}} \exp [i \int_0^\eta \Omega_n
d\eta].\end{equation}

Therefore when $\eta \rightarrow +\infty$ we have

\begin{equation}B_n
\approx -\frac{i}{2}\frac{\dot{g}_m}{g_m}=\frac{1}{2}\Omega_m.
\end{equation}

Then we have two cases:

i) $\Omega_N=\Omega_k$ $\in {\cal R}^+$ for the real factors. Then we see that
when
$\eta \rightarrow +\infty$, the r.h.s. of (\ref{matriz2}) is an oscillatory 
function
with no limit in general. We only have a good limit for some particular
initial conditions \cite{7}(as $Q_m=0$ or $P_m=0$).

ii) $\Omega_m=\Omega_n=E_n-\frac{i}{2}\tau_n^{-1}$ $\in {\cal C}$ for the
complex factors. If we choose the lower Hardy class space $\Phi_-$ to define 
our
rigged Hilbert space we will have a positive imaginary part, and there will 
be a growing factor in the first term of (\ref{gN}) and a damping factor in 
the second
one. In this case, for $a\rightarrow +\infty$, we have a definite limit:

\begin{equation}B_n={1\over{2}}\Omega_n.\label{c4}\end{equation}

>From equations $\left( \text{\ref{potencial}}\right) $, $\left( \text{\ref
{energia
compleja}}\right) $ and $\left( \text{\ref{c4}}\right) $ we can
compute the expression for $B_n$ for both semiclassical solutions 
$\alpha $
and $\beta :$

\begin{eqnarray}
B_n\left( \eta ,\alpha \right) &=&B_n\left( \eta ,\beta \right) =%
{\textstyle {\sqrt{2} \over 4}}
\left[ m^2B^2\eta ^2+\left( m^4B^4\eta ^4+4m^2B^2\left( n+%
{\textstyle {1 \over 2}}
\right) ^2\right) ^{\frac 12}\right] ^{\frac 12}  \label{e1} \\
&&-i\quad \frac{%
{\textstyle {\sqrt{2} \over 2}}
mB\left( n+%
{\textstyle {1 \over 2}}
\right) }{\left[ m^2B^2\eta ^2+\left( m^4B^4\eta ^4+4m^2B^2\left( n+%
{\textstyle {1 \over 2}}
\right) ^2\right) ^{\frac 12}\right] ^{\frac 12}}.  \nonumber
\end{eqnarray}

Now we will see, making the exact calculations, that in the limit 
$\eta
\rightarrow \infty $ there is necessarily decoherence for:

a) different classical geometries ($\alpha $ $\neq $ $\beta 
), $i.e.$ \left|
\rho _r^{\alpha \beta }\left( \eta ,\eta ^{\prime }\right) \right|
\rightarrow 0$ when $\eta \rightarrow \infty $.

b) for the same classical geometry if the times $\eta $ and $\eta 
^{\prime }$
are different, namely $\left| \rho _r^{\alpha \alpha }\left( \eta ,\eta
^{\prime }\right) \right| \rightarrow 0$ and $\left| \rho _r^{\beta \beta
}\left( \eta ,\eta ^{\prime }\right) \right| \rightarrow 0$ when $\eta
\rightarrow \infty .$

>From equations $\left( \text{\ref{agaussiano}}\right) \ $and $\left( 
\text{%
\ref{matriz2}}\right) $ we obtain:

\begin{equation}
\rho _{rn}^{\alpha \beta }\left( \eta ,\eta ^{\prime }\right) =\left( \frac{%
4\,B_{nR}\left( \eta ,\alpha \right) \ B_{nR}\left( \eta ^{\prime },\beta
\right) }{\left[ B_n^{*}\left( \eta ,\alpha \right) +B_n\left( \eta ^{\prime
},\beta \right) \right] ^2}\right) ^{\frac 14}\exp \left[ -i\alpha _n\left(
\eta ,\alpha \right) +i\alpha _n\left( \eta ^{\prime },\beta \right) \right]
.\label{matriz}
\end{equation}

First, we will study decoherence for a) the same semiclassical 
solution but
for different conformal times. Therefore we will calculate the 
asymptotic
behavior $\left( \eta ,\eta ^{\prime }\rightarrow \infty \right) $ of $%
\,\left| \rho _{rn}^{\alpha \alpha }\left( \eta ,\eta ^{\prime }\right)
\right| $, that reads :

\begin{equation}
\left| \rho _{rn}^{\alpha \alpha }\left( \eta ,\eta ^{\prime }\right)
\right| \cong \left[ \frac{4\,\eta \,\eta ^{\prime }}{\left[ \eta +\eta
^{\prime }\right] ^2}\right] ^{\frac 14}.  \label{m}
\end{equation}

Making the following change of variable :\ $\frac{\eta -\eta ^{\prime 
}}2%
=\Delta $ ;\ $\frac{\eta +\eta ^{\prime }}2=\bar{\eta}$ \ with $\Delta 
\ll 1 
$ we obtain:

\begin{equation}
\left| \rho _{rn}^{\alpha \alpha }\left( \eta ,\eta ^{\prime }\right)
\right| \cong \left[ 1-\left( \frac \Delta {\bar{\eta}}\right) ^2\right] ^{%
\frac 14}.  \label{n}
\end{equation}

Since $\left| \rho _{rn}^{\alpha \alpha }\left( \eta ,\eta ^{\prime }\right)
\right| \leq 1$ with the equality only if $\eta =\eta ^{\prime }$, it is
easy to see from Eq. $\left( \text{\ref{matriz2}}\right) $ that $\left| 
\rho
_r^{\alpha \alpha }\left( \eta ,\eta ^{\prime }\right) \right| $ is equal to
zero if $\eta \neq \eta ^{\prime }.$ This means that the reduced 
density
matrix has diagonalized perfectly, i.e. we have achieved perfect
decoherence. However, it is known \cite{6,8if,7} that perfect 
decoherence
also implies that the Wigner function has an infinite spread, so we 
cannot
say that the system is classical.

On the other hand, in Refs. \cite{10,11} working 
with the
consistent histories formalism made the assumption that exactly 
consistent
sets of histories must be found very close to an approximately 
consistent
set. In fact we have found the exact consistent set of histories, so it
would be reasonable to say that there are many approximate 
consistent sets
near of it. Although we are not working with this formalism, we can 
consider
geometries that this statement is also valid in our case. Then, 
having an
exact consistent set of histories means in our formalism exact 
decoherence.
So, we can try to find the approximate decoherence (i.e. the appoximate 
 consistent sets) near the exact one.

\section{Approximate decoherence and classical correlations}

If we introduce a cutoff, $N$ in Eq.$\,\left( \text{\ref{matriz2}}\right) $\ at
some very large value of $n$, the reduced density matrix is 
not diagonal
anymore, i.e. we obtain an approximate decoherence. Let us 
postpone
for the next section the discussion about the value and nature of $N$. Thus we 
obtain
if $\eta \approx \eta ^{\prime }:$%
\begin{equation}
\left| \rho _r^{\alpha \alpha }\left( \eta ,\eta ^{\prime }\right) \right|
=\left| \prod_{n=1}^N\rho _{rn}^{\alpha \alpha }\left( \eta ,\eta ^{\prime
}\right) \right| \approx \exp -\left[ \frac N4\left( \frac \Delta 
{\bar{\eta}%
}\right) ^2\right].   \label{a}
\end{equation}

>From the last equation, we observe that the reduced density matrix 
turns out
to be a gaussian of width $\sigma _d$ where :

\begin{equation}
\sigma _d=\frac{2\ \bar{\eta}}{N^{\frac 12}}.  \label{dec}
\end{equation}

Thus, it must be $\sqrt{N}>>1$ in order to obtain decoherence.

>From equations (\ref{e1}) and (\ref{matriz}) we compute $\left| \rho
_r^{\beta \beta }\left( \eta ,\eta ^{\prime }\right) \right| $ and b) $%
\left| \rho _r^{\alpha \beta }\left( \eta ,\eta ^{\prime }\right) \right| $
and obtain for $\eta \rightarrow \infty $ as in eq. (\ref{m}):

\begin{equation}
\left| \rho _{rn}^{\beta \beta }\left( \eta ,\eta ^{\prime }\right) \right|
=\left| \rho _{rn}^{\alpha \beta }\left( \eta ,\eta ^{\prime }\right)
\right| \cong \left[ \frac{4\,\eta \,\eta ^{\prime }}{\left[ \eta +\eta
^{\prime }\right] ^2}\right] ^{\frac 14}.
\end{equation}

So, following the same steps we did for $\left| \rho _{rn}^{\alpha 
\alpha
}\left( \eta ,\eta ^{\prime }\right) \right| \left[ \text{Eqs. (\ref{m}) to 
(\ref{dec})}\right] $ we can see that the ''decoherence conditions''$\,\left( 
\text{Eq. \ref{dec}}\right) $ are the same for a) case: different conformal 
times,
and b): for different classical geometries. It is easy to see that we 
can
follow the same steps for $\left| \rho _{rn}^{\alpha \beta }\left( \eta
,\eta ^{\prime }\right) \right| $ since from eq. (\ref{e1}) $B_n\left( \eta
,\alpha \right) =B_n\left( \eta ,\beta \right) $.

At this point we will analyze the existence of correlations between 
coordinates and
momenta using Wigner function criterion \cite{9}. Since 
correlations between
coordinates and momenta should be examined ``inside'' each 
classical branch,
we compute Wigner function associated with each
semiclassical solution. The Wigner function associated with the 
reduced density
matrix given by equations $\left( \text{\ref{matriz2}}\right) $ and 
$\left( 
\text{\ref{matriz}}\right) $ is \cite{7}:

\begin{equation}
F_W^{\alpha \alpha }\left( a,P\right) \cong C^2\left( \eta \right) 
\,\sqrt{%
\frac \pi {\sigma _c^2}}\exp \left[ -\frac{\left( P-M^2\dot{S}%
+\sum_{n=1}^N\left( \dot{\alpha}_n-\frac{\dot{B}_{ni}}{4B_{nR}}\right)
\right) ^2}{\sigma _c^2}\right],
\end{equation}
where

\begin{equation}
\sigma _c^2=\sum_{n=1}^N\frac{\left| \dot{B}_n\right| ^2}{4B_{nR}^2}.
\end{equation}

We can predict strong correlation when the centre of the peak of 
Wigner
function is large compared to the spread, i.e., when:

\begin{equation}
\left( M^2\dot{S}-\sum_{n=1}^N\left( \dot{\alpha}_n-
\frac{\dot{B}_{ni}}{%
4B_{nR}}\right) \right) ^2\gg \sigma _c^2.  \label{correlaciones}
\end{equation}

Using the same approximation we made for calculating the reduced 
density
matrix, we obtain the following expression for the width of Wigner 
function:

\begin{equation}
\sigma _c^2\left( \eta ,\alpha \right) \cong \frac N{4\,\eta ^2}
.\label{cor1}
\end{equation}

We can see that the $\sigma _c$ is the inverse of $\sigma _d$ (Eq. 
$\left( 
\text{\ref{a}}\right) $), showing the antagonic relation of 
decoherence and
correlations \cite{7}.

We also calculate the centre of the peak of Wigner function, 
namely:

\begin{equation}
\left( M^2\dot{S}-\sum_{n=1}^N\left( \dot{\alpha}_n-
\frac{\dot{B}_{ni}}{%
4B_{nR}}\right) \right) ^2\cong m^2B^2N^2\eta ^2.  \label{cor2}
\end{equation}

>From equations (\ref{cor1}) and (\ref{cor2}) we it is posible to see the 
behavior 
of the
centre of the peak and the width of Wigner's function in the limit 
$\eta
\rightarrow \infty .$ Thus the condition for the existence of 
correlations
turns out to be:

\begin{equation}
N>>\frac 1{m^2B^2\eta ^4}.  \label{correlaciones2}
\end{equation}

So, if the value of the cutoff is such that $N>>1$ and $N>>\frac 1{%
m^2B^2\eta ^4}$ we can say that the sistem behaves classically: the 
off-diagonal terms of the reduced density matrix are exponentially smaller than
the diagonal terms while we can predict strong correlations between $a\left(
\eta \right) $ and its conjugate momenta.

\subsection{Decoherence and Correlations with a specific value for the
cutoff}

In this subsection we propose and discuss a particular value for the cutoff 
$N$, using a relevant physical scale of the theory, namely, the
Planck scale.

As we already have mentioned, it has been studied that stable and unstable
particles are created in universe expansion\cite{2,5,12}. But,
in this work, we have used only the contribution of the unstable particles
(the poles of the S matrix) to verify the emergence of the classical
behavior. Thus, a reasonable choice for the value of $N$ might be to
consider in Eq. $\left( \text{\ref{a}}\right) $ only those unstable
particles (poles) whose mean life is bigger than Planck's time ($t_p=M^{-1}$
in our units). This implies that particles with smaller life time will be 
considered to be outside the domain of our semiclassical quantum gravity model.

In order to calculate the mean life of each pole we have to transform
equations $\left( \text{\ref{energia compleja}}\right) $, $\left( \text{\ref
{vida
media}}\right) \ $and $\left( \text{\ref{e1}}\right) $ to the
non-rescaled case, namely the physical energy is $\frac{\Omega _n}a$ and the
physical decaying time is $\tau _n^{\prime }=a\tau _n.$ Thus from $\
\left( \text{\ref{vida media}}\right) $ we obtain for $\eta \rightarrow
\infty $ the mean life of the unstable state n:

\begin{equation}
\tau _n^{\prime }=\frac{B\,\eta _{out}^2}{\left( n+\frac 12\right) }.
\end{equation}

Thus, with this choice, we consider in Eq. $\left( \text{\ref{a}}\right) $
only those unstable particles with mean life:

\begin{equation}
\tau _n^{\prime }=\frac{B\,\eta ^2}{\left( n+\frac 12\right) }>\frac 1M=t_p
.\end{equation}

Therefore the value of the cutoff turns out to be $N\,=\,M\ B\ \eta ^2$. 
It could be argued that this particar value of $N$ depends of the conformal
time $\eta ,$ but it should be noted that $\frac N{a^2\left( \eta \right) }$
does not depend on $\eta $ anymore. Therefore, $N=N\left( \eta 
\right) $
should be regarded as a consequence of the universe 
expansion. The
reduced density matrix (Eq. (\ref{a})) turns out to be a Gaussian of 
width $%
\sigma _d$ where:

\begin{equation}
\sigma _d=\frac{2\ \eta }{N^{\frac 12}}=\frac 2{\left( M\,B\right) ^{\frac 
12%
}};  \label{sigma}
\end{equation} and, as $\eta =\left( \frac{2\ t}B\right) ^{\frac 12}$, we 
obtain the following expression for the ratio $%
{\displaystyle {\sigma \over \eta}}
$ as a function of t.

\begin{equation}
\frac{\sigma _d}\eta =\sqrt{\frac 2{M\ t}}\approx \sqrt{\frac{\ t_p}t}  
.\label{dec2}
\end{equation}

Therefore the off-diagonal terms will be exponentially smaller than 
the
diagonal terms for $t\ >>\ 
{\displaystyle {1 \over M}}
=t_p.$

With $N=M\,B\,\eta ^2$,we obtain the following expression for eq. 
$\left( 
\text{\ref{correlaciones}}\right) $:

\begin{equation}
m^2M\,B^3\eta ^6>>1
.\end{equation}

Writing the last equation as a function of the physical time $t$, we 
obtain
the condition for the existence of strong correlations :

\begin{equation}
t>>\left( \frac{t_p\ }{8\ m^2\ }\right) ^{\frac 13}
.\end{equation}

\section{Conclusions}

We have shown that the S-matrix of a quantum 
field theory in
curved space model has an infinite set of poles. The presence of
these singularities produce the appearance of unstable ideal generalized 
states (with 
complex eigenvalues) in the Universe evolution. The corresponding eigenvectors 
are Gamow vectors and produce exponentially decaying terms. The best feature 
of these decaying terms is 
that they simplify and clarify calculations. The Universe expansion 
leads to decoherence if this expansion produces 
particles creation as well. Our unstable states enlarge the set of 
initial conditions where we can prove that decoherence occurs. In fact, the 
damping factors allow that the interference elements of the reduced 
density matrix dissapear for almost any non-equilibrium initial 
condition of the matter fields. Following the standard procedures, we have 
also shown that the unstable ideal generalized states satisfy the correlation 
 conditions, which, with the decoherence phenomenon, are the origin of 
the semiclassical Einstein equations.

The conditions about decoherence and correlations were imposed by 
means of an ultraviolet cutoff, $N$, related with the energy scale 
where the semiclassical approximation is taken as valid. The introduction 
of this cutoff in relevant in order to preserve both necesary conditions for 
calssicality: decoherence plus correlations. Without the presence of the 
cutoff the infinite set of unstable codes destroy the classical correlattion 
and the semiclassical limit would be untanable.

Decoherence is the key to understanding the relationship between the arrows
of time in cosmology.  In the context of quantum open systems, where the
metric is viewed as the ``system'' and the quantum fields as the
``environment,'' decoherence is produced by the continuous interaction between
system and environment.  The non-symmetric transfer of information from system
to environment is the origin of an entropy increase (in the sense of von
Neumann), because there is loss of information in the system, and of the time
asymmetry in cosmology, because growth of entropy, particle creation and
isotropization show a tendency towards equilibrium.  However, decoherence is
also a necessary condition for the quantum to classical transition.  In the
density matrix formulation, decoherence appears as the destruction of
interference terms and, in our model, as the transition from a pure to a mixed
state in the time evolution of the density matrix associated with the RW
metric; the interaction with the quantum modes of the scalar fields is the
origin of such a non-unitary evolution.

It is interesting to note that, in the cosmological model we considered, 
unstable particle
creation and decoherence are the effect of resonances between the evolutions
of the scale factor $a$ and the free massive field, which is, on 
the other hand, the origin
of the chaotic behaviour in the classical evolution of the cosmological model
\cite{fer1}. This observation opens a new and interesting path in the study 
of the relationship between classical chaotic models and the decoherence 
phenomena.

\section*{Acknowledgments}

This work was supported by Universidad de Buenos Aires, CONICET and
Fundaci\'on Antorchas.

\end{document}